\documentstyle[12pt,axodraw,epsf]{article} 
\begin{document}

\begin{flushright}
   CERN-TH/99-379\\[-0.2cm]
   THES-TP/99-14\\[-0.2cm]
   hep-ph/9912253\\[-0.2cm]
   December 1999
\end{flushright}

\begin{center}
{\Large {\bf Loop-Induced CP Violation in the Gaugino and\\[0.3cm]
Higgsino Sectors of Supersymmetric Theories}}\\[1.4cm]
{\large Apostolos Pilaftsis}\\[0.4cm]
{\em Theory Division, CERN, CH-1211 Geneva 23, Switzerland\\[0.2cm]
and\\[0.2cm]
Department of Theoretical Physics, University of Thessaloniki,\\
GR 54006 Thessaloniki, Greece}
\end{center}
\bigskip\bigskip \centerline{\bf  ABSTRACT}  
We  show that  the  gaugino  and higgsino  sectors of   supersymmetric
theories  can  naturally  acquire   observable   CP  violation through
radiative  effects which  originate  from large CP-violating trilinear
couplings  of the Higgs bosons to  the third-generation scalar quarks.
These  CP-violating loop  effects are not  attainable  by evolving the
supersymmetric     renormalization-group  equations    from   a higher
unification scale down to the electroweak one.  We briefly discuss the
phenomenological consequences of  such a scenario,  and as an example,
calculate the  two-loop  contribution to the  neutron  electric dipole
moment generated by the one-loop chromo-electric  dipole moment of the
gluino.\\[0.25cm] 
PACS numbers: 11.30.Er, 14.80.Er

\newpage

Supersymmetric (SUSY)  theories, including the  minimal supersymmetric
Standard Model  (MSSM), predict several  new unsuppressed CP-violating
phases which generally lead to too large contributions to the electric
dipole    moments    (EDM's)    of    the   neutron    and    electron
\cite{EFN,DGH,DDLPD,PN,IN,CKP}. Several suggestions  have been made in
the literature to  overcome such a CP crisis  in SUSY theories.  Apart
from the cancellation mechanism  proposed in \cite{IN}, an interesting
solution to  the above CP-crisis problem  is to assume  that the first
two generations are either very  high above the TeV scale \cite{PN,GD}
or  they  do  not  involve  CP-violating  phases  in  their  trilinear
couplings $A_f$  to the Higgs  bosons \cite{CKP}.  In the  latter SUSY
framework, with  all scalar quarks much  below the TeV  scale, we also
have  to require that  the SU(3)$_c$,  SU(2)$_L$ and  U(1)$_Y$ gaugino
masses: $m_{\tilde{g}}$, $m_{\tilde{W}}$  and $m_{\tilde{B}}$, as well
as the higgsino-mass term $\mu$  be real parameters. Then, for both of
the  above scenarios,  only  the scalar-top  ($\tilde{t}$) and  bottom
quarks ($\tilde{b}$)  may be considered  as potential mediators  of CP
non-conservation.\footnote[1]{Nevertheless,   it    might   still   be
  necessary to  assume that CP violation in  the $K^0\bar{K}^0$ system
  is described by  the usual Cabbibo-Kobayashi-Maskawa matrix.}  Here,
we shall adopt  this minimal model of SUSY CP  violation.  We will not
address the issue concerning the underlying mechanism of SUSY breaking
that leads to the above low-energy scenario, as it is beyond the scope
of the present study (see also \cite{GD}).

In the above  discussion, the CP-violating phases were  defined in the
weak  basis in  which the  soft-SUSY-breaking parameter  $m^2_{12}$ of
Higgs mixing is  real.  At the tree level, such  a convention leads to
positive vacuum expectation  values $v_1$ and $v_2$ for  the two Higgs
doublets $\Phi_1$ and  $\Phi_2$ of the MSSM, where  the relative phase
$\xi$ between $v_1$  and $v_2$ vanishes.  If this  phase convention is
to be adopted order by  order in perturbation theory \cite{APLB}, then
the formally radiatively-induced phase  $\xi$ can always be eliminated
by  an appropriate  choice  of the  renormalization counter-term  (CT)
${\rm Im}\,  m^2_{12}$.  This latter  CT is related to  CP-odd tadpole
graphs of the pseudoscalar Higgs bosons, and is important in rendering
the  CP-violating Higgs  scalar-pseudoscalar  transitions ultra-violet
finite   \cite{APLB}    in   Higgs-boson   mass    spectrum   analyses
\cite{PW,CEPW}.   However, it  should  be stressed  that unlike  ${\rm
  Re}\, m^2_{12}$,  ${\rm Im}\, m^2_{12}$ does not  directly enter the
renormalization of the physical kinematic parameters of the MSSM, such
as  the   charged  Higgs-boson  mass   and  $\tan\beta  =   v_2/v_1$.  
Consequently, no additional CP  violation can be generated through the
phase $\xi$ in the effective chargino and neutralino mass matrices.

In  this  paper, we  shall  study  a  novel mechanism  of  radiatively
inducing CP violation in the gaugino and higgsino sectors of the MSSM.
In  these   sectors,  CP   violation  is  communicated   by  trilinear
interactions of scalar top and  bottom quarks to the Higgs bosons.  It
is interesting to remark that this loop-induced CP violation cannot be
achieved by  running the  renormalization-group (RG) equations  from a
higher unification  scale down to the electroweak  energies \cite{GW}. 
Even though the discussion is confined to the MSSM, the results of the
analysis can straightforwardly be generalized to other SUSY extensions
of the SM.

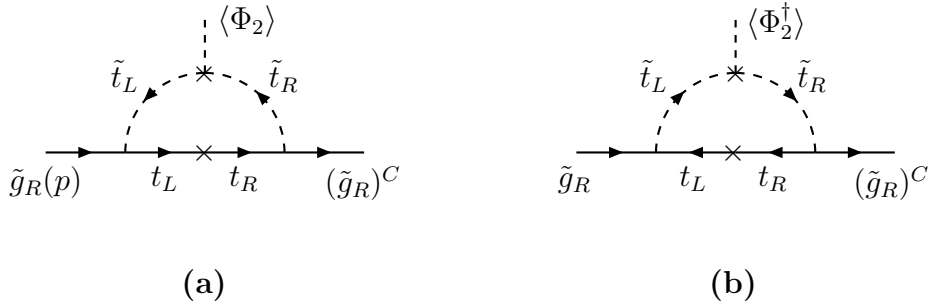
\begin{figure}[t]

\begin{center}
\begin{picture}(300,100)(0,0)
\SetWidth{0.8}
 
\ArrowLine(0,50)(30,50)\ArrowLine(30,50)(60,50)
\ArrowLine(60,50)(90,50)\ArrowLine(90,50)(120,50)
\DashArrowArc(60,50)(30,0,90){3}\DashArrowArc(60,50)(30,90,180){3}
\DashLine(60,80)(60,100){3}\Text(65,100)[l]{$\big< \Phi_2 \big>$}
\Text(60,80)[]{\bf $\times$}\Text(60,50)[]{\bf $\times$}
\Text(0,45)[t]{$\tilde{g}_R (p)$}\Text(120,45)[t]{$(\tilde{g}_R)^C$}
\Text(45,45)[t]{$t_L$}\Text(75,45)[t]{$t_R$}
\Text(30,75)[b]{$\tilde{t}_L$}\Text(90,75)[b]{$\tilde{t}_R$}

\Text(60,0)[]{\bf (a)}

\ArrowLine(200,50)(230,50)\ArrowLine(260,50)(230,50)
\ArrowLine(290,50)(260,50)\ArrowLine(290,50)(320,50)
\DashArrowArcn(260,50)(30,180,90){3}\DashArrowArcn(260,50)(30,90,0){3}
\DashLine(260,80)(260,100){3}\Text(265,100)[l]{$\big< \Phi_2^\dagger \big>$}
\Text(261,80)[]{\bf $\times$}\Text(260,50)[]{\bf $\times$}
\Text(200,45)[t]{$\tilde{g}_R$}\Text(320,45)[t]{$(\tilde{g}_R)^C$}
\Text(245,45)[t]{$t_L$}\Text(275,45)[t]{$t_R$}
\Text(230,75)[b]{$\tilde{t}_L$}\Text(290,75)[b]{$\tilde{t}_R$}

\Text(260,0)[]{\bf (b)}

\end{picture}
\end{center}
\caption{Chromo-electric  dipole moment  of  the gluino; attaching the
gluon field to the top- and scalar-top- quark lines is implied.}\label{f1}
\end{figure}

\begin{figure}[t]

\begin{center}
\begin{picture}(100,100)(0,0)
\SetWidth{0.8}
 
\ArrowLine(0,50)(30,50)\ArrowLine(30,50)(60,50)
\ArrowLine(60,50)(90,50)\ArrowLine(90,50)(120,50)
\DashArrowArcn(60,50)(30,90,0){3}\DashArrowArcn(60,50)(30,180,90){3}
\DashLine(60,80)(60,100){3}\Text(65,100)[l]{$\big< \Phi_{1,2} \big>$}
\Text(61,80)[]{\bf $\times$}\GCirc(60,50){5}{0.7}
\Text(0,45)[t]{$q_L$}\Text(120,45)[t]{$q_R$}
\Text(45,45)[t]{$\tilde{g}_R$}\Text(80,45)[t]{$\tilde{g}_L$}
\Text(30,75)[b]{$\tilde{q}_L$}\Text(90,75)[b]{$\tilde{q}_R$}
\Gluon(60,45)(60,15){2}{3}\Text(65,15)[l]{$g_\mu (k)$}

\end{picture}
\end{center}
\caption{Two-loop CEDM of  a light quark induced  by the one-loop CEDM
of the gluino.}\label{f2}
\end{figure}
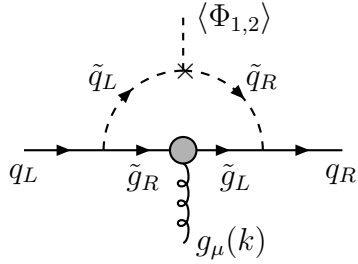

We start   our discussion by   considering the gluino sector.   In the
convention   in which  the gluino  mass   is real,  i.e.\ ${\rm arg}\,
(m_{\tilde{g}} )   = 0$, CP   violation is induced  by the interaction
Lagrangian:
\begin{equation}
  \label{gqq}
{\cal L}^{\tilde{g}\tilde{q}q}_{\rm int}\ =\ 
-\sqrt{2} g_s\, \sum\limits_{q=t,b} \bigg(\: \tilde{q}^*_L\, 
\bar{\tilde{g}}^a\,
\frac{\lambda^a}{2}\,P_L q\: -\: \tilde{q}^*_R\, \bar{\tilde{g}}^a\,
\frac{\lambda^a}{2}\, P_R q\: \bigg)\: +\: {\rm h.c.},
\end{equation}
where $P_{L (R)}  = (1 -(+)\gamma_5 )/2$ is  the left- (right-) handed
chirality projection operator   and $\lambda^a$, with $a=1,\dots  ,8$,
are the usual SU$(3)$ Gell-Mann matrices. In fact, CP violation enters
through the mixing of the  weak $(\tilde{q}_L, \tilde{q}_R)$ with mass
eigenstates $(\tilde{q}_1, \tilde{q}_2)$.  Such a mixing  of states is
described by the unitary transformation:
\begin{equation}
  \label{Rscalar}
\left( \begin{array}{c} \tilde{q}_L \\ \tilde{q}_R \end{array} \right)
\ =\ U^q\, \left( \begin{array}{c} \tilde{q}_1 \\ \tilde{q}_2
\end{array}\right)\ =\
\left( \begin{array}{cc} 1 & 0 \\ 0 & e^{i\delta_q} \end{array} \right)
\left( \begin{array}{cc} \cos\theta_q & \sin\theta_q \\
          -\sin\theta_q & \cos\theta_q \end{array} \right)\,
\left( \begin{array}{c} \tilde{q}_1 \\ \tilde{q}_2 \end{array}\right)\, ,
\end{equation}
where  $\delta_q  =  {\rm  arg}\,  (A_q -  \mu^*  R_q)$,  with  $R_t =
\cot\beta$ and $R_b = \tan\beta$, is a non-trivial CP-violating phase.
The gluino mass $m_{\tilde{g}}$ receives a finite radiative shift from
diagrams depicted in Fig.\ \ref{f1}, which can easily be calculated to
be
\begin{equation}
  \label{dmg}
\delta m_{\tilde{g}}\ =\ \frac{\alpha_s}{2\pi}\, m_t \sin\theta_t  
\cos\theta_t\, e^{-i\delta_t}\, \int\limits_0^1 dx\, \ln\bigg[\,
\frac{m^2_t x\, +\, M^2_{\tilde{t}_1} (1-x)}{
m^2_t x\, +\, M^2_{\tilde{t}_2} (1-x)}\,\bigg]\, .
\end{equation}
It is  obvious that  only scalar-top quarks  are of relevance,  as the
effect  of  the  scalar  bottom  quarks  is  suppressed  by  a  factor
$m^2_b/m^2_t$.  For relatively large  values of the scalar-top mixing,
i.e.\  $M_{\tilde{t}_2}/  M_{\tilde{t}_1}  \sim  5$,  $\theta_t\approx
45^\circ$,  we find a  radiatively-induced phase  for $m_{\tilde{g}}$:
${\rm arg}  (m_{\tilde{g}}) \approx 10^{-2}  m_t/m_{\tilde{g}}$.  Even
though ${\rm  arg} (m_{\tilde{g}})$  may formally be  re-absorbed into
the  renormalization of $m_{\tilde{g}}$,  its presence  will, however,
manifest  itself  in  related  CP-violating  higher-point  correlation
functions,\footnote[2]{As  was  discussed  in \cite{DGH},  ${\rm  arg}
  (m_{\tilde{g}})$ may  also contribute  to the CP-violating  phase of
  QCD  $\bar{\theta}$.  Here  we shall  not deal  with this  strong CP
  problem, but  merely assume that  $\bar{\theta}$ is eliminated  by a
  chiral rotation  of the quark fields.}  such  as the chromo-electric
dipole  moment (CEDM)  form  factor  of the  gluino  described by  the
effective  Lagrangian ${\cal  L}^{\tilde{g}}_{\rm CEDM}  = \frac{1}{4}
d^{\tilde{g}}\,  f^{abc}  \bar{\tilde{g}}^b i\sigma^{\mu\nu}  \gamma_5
\tilde{g}^c  F^a_{\mu\nu}$, where  $F^a_{\mu\nu}$ is  the  gluon field
strength tensor and
\begin{equation}
  \label{CEDMg}
\frac{d^{\tilde{g}}(p^2)}{g_s}\ =\ \frac{\alpha_s}{4\pi}\ m_t\, 
\sum\limits_{i=1,2}\, {\rm Im}\, (U^{t*}_{1i} U^t_{2i})\, 
\int\limits_0^1 dx\ \frac{x}{m^2_t x\, +\, M^2_{\tilde{t}_i} (1-x)\, 
-\, p^2 x(1-x) }\ .
\end{equation}
Notice that the gluino being a  Majorana particle with internal degrees of
freedom,  i.e.\ colour,  can only   possess  a non-vanishing  CEDM; it
cannot have an EDM.  Furthermore,  at two loops,  the CEDM form factor
of Eq.\ (\ref{CEDMg}) can in turn give rise to a CEDM of a light quark
$d^C_q$, as  shown in Fig.\  \ref{f2}. The  contribution to CEDM  of a
light quark $d^C_q$ has been calculated to give
\begin{equation}
  \label{edm2g}
\frac{d^C_q}{g_s}\ =\ \frac{3\alpha^2_s}{16\pi^2}\, 
\frac{m_t}{m^2_{\tilde{g}}}\,
\sum\limits_{i,j=1,2}\, {\rm Im}\, (U^{t*}_{1i} U^t_{2i})\
{\rm Re}\, (U^{q*}_{1j} U^q_{2j})\
H\bigg(\frac{M^2_{\tilde{q}_j}}{m^2_{\tilde{g}}}\, , 
\frac{m^2_t}{m^2_{\tilde{g}}}\, ,
\frac{M^2_{\tilde{t}_i}}{m^2_{\tilde{g}}}\bigg)\, ,
\end{equation}
where the two-loop function $H(a,b,c)$ is defined as
\begin{eqnarray}
  \label{HH}
H(a,b,c) &=& \int_0^1 \frac{dx\, x}{(1-a)^2}\, \bigg[\, \frac{1-a}{
x(1-x) - bx - c (1-x)}\: +\: \frac{a\ln a}{ax(1-x) - bx - c (1-x)}\nonumber\\
&&\hspace{-1.6cm}
-\, \frac{(1-a)^2\, [bx + c(1-x)]\,x(1-x)}{[ax(1-x) - bx - c (1-x)]
[x(1-x) - bx - c (1-x)]^2}\, \ln\bigg(\frac{bx + c (1-x)}{
x(1-x)}\bigg)\, \bigg]\, .
\end{eqnarray}
Based on the naive quark model and after including QCD renormalization
effects, the contribution to the neutron EDM may be estimated to be 
\begin{equation}
  \label{dne}
\frac{d^C_n}{e}\ \approx\ 
\bigg(\,\frac{g_s (M_Z )}{g_s (\Lambda )}\,\bigg)^{120/23}\
\bigg[\, \frac{4}{9}\,\bigg(\frac{d^C_d}{g_s}\bigg)_\Lambda\ +\ 
\frac{2}{9}\,\bigg(\,\frac{d^C_u}{g_s}\,\bigg)_\Lambda
\bigg]\, ,
\end{equation}
where $g_s$ and  the $u$- and $d$-quark masses  in $d^C_u$ and $d^C_d$
are calculated at the scale $\Lambda$:  $m_u (\Lambda ) = 7$ MeV, $m_d
(\Lambda  )  = 10$  MeV,  and  $g_s  (\Lambda )/(4\pi)  =  1/\sqrt{6}$
\cite{DDLPD}. 


\begin{table}[t]

\begin{center}

\begin{tabular}{|cc||ccc|| ccc|}
\hline
$\tan\beta$ & 
$\widetilde{M}_{u,d}$ & & $|d^C_n/e|$ & $[\,10^{-26}\ {\rm cm}\,]$ &
& $|d^C_n/e|$ & $[\,10^{-26}\ {\rm cm}\,]$ \\
& $[\,{\rm GeV}\,]$ & 
$m_{\tilde{g}}$~~~~$=$ & $200$~~GeV & & $m_{\tilde{g}}$~~~~$=$
& $400$~~GeV &\\
& & $\mu =$ 0.5, & ~~~~~1,~~~~~ & 2~TeV & $\mu =$ 0.5, 
& ~~~~~1,~~~~~ & 2~TeV\\ 
\hline\hline
   & 150 & 0.3 & 1.0 & 5.0 & 0.2 & 0.6 & 3.0 \\
2  & 300 & 0.1 & 0.2 & 1.1 & 0.1 & 0.2 & 0.9 \\
\hline
   & 150 & 1.3 & 4.2 & 10. & 0.8 & 2.7 & 6.6 \\
5  & 300 & 0.3 & 0.9 & 2.3 & 0.3 & 0.8 & 2.0 \\
\hline
   & 150 & 9.7 & 21. & 44. & 6.2 & 13. & 28. \\
20 & 300 & 2.2 & 4.7 & 9.8 & 1.9 & 4.1 & 8.4 \\
   & 500 & 0.5 & 1.2 & 2.4 & 0.6 & 1.2 & 2.6 \\
\hline
\end{tabular}
\end{center}

\caption{Numerical estimates of the gluino-induced CEDM contribution
to the neutron EDM, assuming moduli-universal trilinear couplings:
$|A_f| = 1$ TeV, with ${\rm arg} (A_{t,b}) = 90^\circ$ and ${\rm arg}
(A_{u,d}) = 0$, and non-universal soft-scalar-quark masses:
$\widetilde{M}_{t_L} = \widetilde{M}_{t_R} = 500$ GeV and
$\widetilde{M}_{u,d} = \widetilde{M}_{u_L,d_L} =
\widetilde{M}_{u_R,d_R}$ as given in the table.}\label{tab1}
\end{table}

In Table \ref{tab1}, we  exhibit numerical estimates of $d^C_n/e$, for
a    non-universal  scenario   for     the  scalar-quark  masses  with
moduli-universal  trilinear couplings.  We  assume that  the first two
generations  possess  no CP  violation,  and the only non-vanishing CP
phase  is ${\rm  arg}  (A_{t,b}) =  90^\circ$.   We find that  the EDM
contributions increase  almost   linearly  with  $\tan\beta$  and,  as
expected,   decouple  when  the  first-generation scalar-quark  masses
$\widetilde{M}_{u,d}$ become large.   The additional EDM contributions
to  $d_n$ are  comparable to its    present experimental upper  limit:
$|d_n| <  6.3\times 10^{-26}$  \cite{EDMexp}, for $\widetilde{M}_{u,d}
\stackrel{<}{{}_\sim} 300$    GeV,  and relatively   large  scalar-top
mixing.  In particular, it would be interesting to remark that for the
same  values of the  $\mu$-term and the soft-SUSY-breaking parameters,
the two-loop EDM effects induced by the  gluino CEDM are comparable in
size to the two-loop EDM  contributions due to Weinberg's  three-gluon
operator,  for     $m_{\tilde{g}}   \stackrel{>}{{}_\sim}  400$    GeV
\cite{DDLPD,IN},  and   to   the  two-loop  Barr-Zee-type  EDM  graphs
\cite{CKP}, for `CP-odd' scalar masses larger than 700 GeV.

\begin{figure}[t]

\begin{center}
\begin{picture}(300,200)(0,0)
\SetWidth{0.8}
 
\ArrowLine(0,150)(30,150)\ArrowLine(30,150)(60,150)
\ArrowLine(60,150)(90,150)\ArrowLine(90,150)(120,150)
\DashArrowArc(60,150)(30,0,90){3}\DashArrowArc(60,150)(30,90,180){3}
\DashLine(60,180)(60,200){3}\Text(65,200)[l]{$\big< \Phi_1 \big>$}
\Text(60,180)[]{\bf $\times$}\Text(60,150)[]{\bf $\times$}
\Text(0,145)[t]{$\bar{\tilde{h}}{}^-_1 (p)$}\Text(120,145)[t]{$\tilde{h}^+_2$}
\Text(45,145)[t]{$t_L$}\Text(75,145)[t]{$t_R$}
\Text(30,175)[b]{$\tilde{b}_R$}\Text(90,175)[b]{$\tilde{b}_L$}

\Text(60,120)[]{\bf (a)}

\ArrowLine(200,150)(230,150)\ArrowLine(230,150)(260,150)
\ArrowLine(260,150)(290,150)\ArrowLine(290,150)(320,150)
\DashArrowArc(260,150)(30,90,180){3}\DashArrowArc(260,150)(30,0,90){3}
\DashLine(260,180)(260,200){3}\Text(265,200)[l]{$\big< \Phi_2 \big>$}
\Text(261,180)[]{\bf $\times$}\Text(260,150)[]{\bf $\times$}
\Text(200,145)[t]{$\tilde{h}^-_1$}\Text(320,145)[t]{$\bar{\tilde{h}}{}^+_2$}
\Text(245,145)[t]{$b_R$}\Text(275,145)[t]{$b_L$}
\Text(230,175)[b]{$\tilde{t}_L$}\Text(290,175)[b]{$\tilde{t}_R$}

\Text(260,120)[]{\bf (b)}

\ArrowLine(0,50)(30,50)\ArrowLine(30,50)(90,50)\ArrowLine(90,50)(120,50)
\DashArrowArc(60,50)(30,0,90){3}\DashArrowArc(60,50)(30,90,180){3}
\DashLine(60,80)(60,100){3}\Text(65,100)[l]{$\big< \Phi_2 \big>$}
\Text(60,80)[]{\bf $\times$}
\Text(0,45)[t]{$\bar{\tilde{h}}{}^+_2$}
\Text(120,45)[t]{$\bar{\tilde{w}}{}^+$}
\Text(60,45)[t]{$b_L$}
\Text(30,75)[b]{$\tilde{t}_R$}\Text(90,75)[b]{$\tilde{t}_L$}

\Text(60,20)[]{\bf (c)}

\ArrowLine(200,50)(230,50)\ArrowLine(230,50)(290,50)\ArrowLine(290,50)(320,50)
\DashArrowArc(260,50)(30,0,90){3}\DashArrowArc(260,50)(30,90,180){3}
\DashLine(260,80)(260,100){3}\Text(265,100)[l]{$\big< \Phi_1 \big>$}
\Text(261,80)[]{\bf $\times$}
\Text(200,45)[t]{$\bar{\tilde{h}}{}^-_1$}
\Text(320,45)[t]{$\bar{\tilde{w}}{}^-$}
\Text(260,45)[t]{$t_L$}
\Text(230,75)[b]{$\tilde{b}_R$}\Text(290,75)[b]{$\tilde{b}_L$}

\Text(260,20)[]{\bf (d)}

\end{picture}
\end{center}
\caption{CP-violating    self-energy   transitions  in  the   chargino
sector.}\label{f3}
\end{figure}
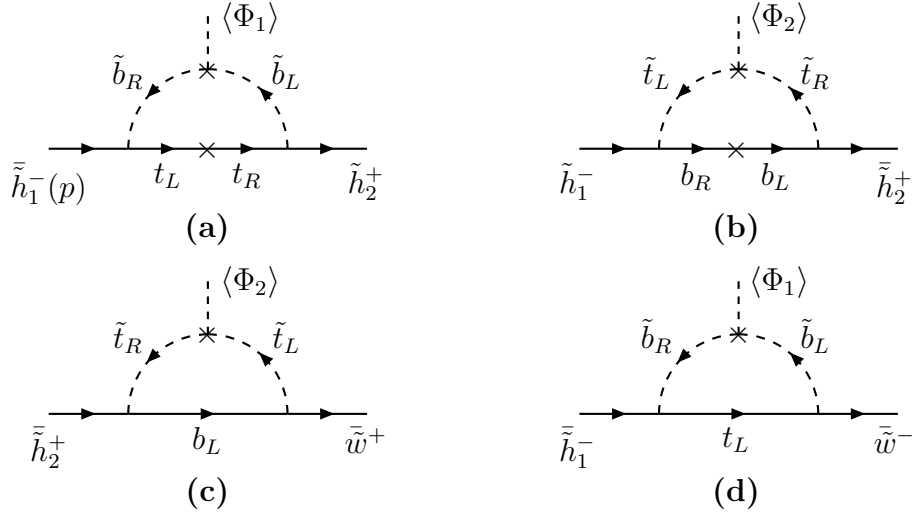

In the following, we will focus  our attention on the chargino sector. 
Significant  CP  violation  in  this  sector can  be  induced  by  the
interactions of the  charged gaugino, $\widetilde{W}^T = (\tilde{w}^+,
\bar{\tilde{w}}{}^-)$,     and     higgsino,    $\widetilde{H}^T     =
(\tilde{h}^+_2, \bar{\tilde{h}}{}^-_1)$  with $\tilde{t}$, $\tilde{b}$
and $t$, $b$.  Specifically, the interaction Lagrangian of interest to
us reads:
\begin{eqnarray}
  \label{chqq}
{\cal L}^{(\widetilde{W},\widetilde{H})\tilde{q}q}_{\rm int} &=&
-g_w\, \Big(\, \tilde{b}^*_L\, \overline{\widetilde{W}} P_L t\: +\:
\tilde{t}^*_L\, \overline{\widetilde{W}}{}^C\! P_L b\, \Big)\: +\:
h_b\,\Big(\, \tilde{b}^*_R\,\overline{\widetilde{H}} P_L t\: +\:
\tilde{t}^*_L\, \overline{\widetilde{H}}{}^C\! P_R b\, \Big)\nonumber\\
&&+\, h_t\, \Big(\, \tilde{b}^*_L\,\overline{\widetilde{H}} P_R t\: +\:
\tilde{t}^*_R\, \overline{\widetilde{H}}{}^C\! P_L b\, \Big)\ +\
{\rm h.c.}
\end{eqnarray}
Again,  the  graphs   shown  in  Fig.\  \ref{f3}  can   give  rise  to
CP-violating  self-energy transitions among  the chargino  fields.  In
particular,  Fig.\  \ref{f3}(c),   which  represents  an  off-diagonal
wave-function   contribution   to   the   $\widetilde{W}\widetilde{H}$
transition,  becomes  dominant; the  other  graphs  are suppressed  by
powers of $m_b/m_t$.  To have an estimate of CP violation, we consider
the effective kinetic Lagrangian
\begin{equation}
  \label{LCPkin}
{\cal L}^{\rm kin}_{\rm eff}\ =\ 
\Big(\overline{\widetilde{W}},\ \overline{\widetilde{H}}\Big)
\left(\! \begin{array}{lr}
\not\! p - m_{\widetilde{W}} &\hspace{-2.55cm}
\Sigma_L\! \not\! p P_L -
\frac{1}{\sqrt{2}}\,g_w v_1 P_R - \frac{1}{\sqrt{2}}\,g_w v_2 P_L \\
\Sigma^*_L\! \not\! p P_L -
\frac{1}{\sqrt{2}}\,g_w v_2 P_R - \frac{1}{\sqrt{2}}\,g_w v_1 P_L
& \not\! p - \mu\end{array}\! \right)
\left(\!\begin{array}{c}\widetilde{W} \\
\widetilde{H}\end{array}\!\right),
\end{equation}
with 
\begin{equation}
  \label{SigmaL} 
\Sigma_L (p^2=0)\   =\ \frac{3  g_w   h_t}{32\pi^2}\, \sin\theta_t
\cos\theta_t\, e^{-i\delta_t}\,
\ln\bigg(\frac{M^2_{\tilde{t}_1}}{M^2_{\tilde{t}_2}}\bigg)\, .
\end{equation}
Notice  that $\Sigma_L$  should not  be renormalized  away,  since the
respective  $\widetilde{W}\widetilde{H}$-wave-function  CT would  have
hardly   violated  the   original  R-symmetry   of  the   MSSM   by  a
dimension-four operator;  the R-symmetry is only softly  broken in the
MSSM by the  gaugino masses, the trilinear couplings  $A_f$ and $\mu$. 
Under the R-symmetry, matter superfields and gauginos carry R-charges,
while  Higgs superfields are  neutral.  After  canonically normalizing
the derivative part  of ${\cal L}^{\rm kin}_{\rm eff}$,  we obtain the
effective CP-violating chargino mass matrix:
\begin{equation}
  \label{Mchargino}
{\cal L}^{\rm mass}_{\rm  eff}\ \approx\ \Big(\overline{\widetilde{W}}_R,\
\overline{\widetilde{H}}_R\Big) \left(\! \begin{array}{cc}
m_{\widetilde{W}} - \frac{1}{2\sqrt{2}} g_w v_2 \Sigma^*_L &
\frac{1}{\sqrt{2}} g_w v_2 - \frac{1}{2} m_{\widetilde{W}} \Sigma_L\\
\frac{1}{\sqrt{2}} g_w v_1 - \frac{1}{2}\mu \Sigma^*_L & 
\mu - \frac{1}{2\sqrt{2}} g_w v_1 \Sigma_L \end{array}\!\right) 
\left(\!\begin{array}{c}\widetilde{W}_L \\
\widetilde{H}_L \end{array}\!\right)\ +\ {\rm h.c.}
\end{equation}
{}From  Eq.\ (\ref{Mchargino}),  it is  not difficult  to see  that CP
violation becomes  of order  unity in the  chargino sector,  when $v_1
\sim  \mu {\rm  Im}\, \Sigma_L$.   This last  condition can  be easily
satisfied for  relatively large  scalar-top mixing.  For  example, for
$M_{\tilde{t}_2} / M_{\tilde{t}_1} \sim 3$  and $\mu,\ A_t \sim 1$ TeV
($\theta_t  \approx  45^\circ$), we  have  ${\rm  Im}\, \Sigma_L  \sim
10^{-2}$ and  $\mu {\rm Im}\, \Sigma_L  \sim 10$ GeV,  which is indeed
comparable to $v_1$ for $\tan\beta > 2$.

\begin{figure}[t]

\begin{center}
\begin{picture}(300,100)(0,0)
\SetWidth{0.8}
 
\ArrowLine(0,50)(20,50)\ArrowLine(20,50)(40,50)
\ArrowLine(40,50)(60,50)\ArrowLine(60,50)(100,50)
\ArrowLine(100,50)(120,50)\ArrowLine(120,50)(140,50)
\DashArrowArcn(70,50)(50,180,0){3}
\DashArrowArc(80,50)(20,0,90){3}\DashArrowArc(80,50)(20,90,180){3}
\Text(40,50)[]{\bf $\times$}\Text(80,70)[]{\bf $\times$}
\Text(0,45)[t]{$e_R$}\Text(140,45)[t]{$e_L$}
\Text(30,45)[t]{$\tilde{h}^-_1$}\Text(50,45)[t]{$\bar{\tilde{h}}{}^+_2$}
\Text(110,45)[t]{$\bar{\tilde{w}}{}^+$}
\Text(60,67)[b]{$\tilde{t}_R$}\Text(100,67)[b]{$\tilde{t}_L$}
\Text(80,45)[t]{$b_L$}\Text(60,105)[b]{$\tilde{\nu}_{eL}$}

\Text(60,0)[]{\bf (a)}

\ArrowLine(200,50)(220,50)\ArrowLine(220,50)(240,50)
\ArrowLine(240,50)(260,50)\ArrowLine(260,50)(300,50)
\ArrowLine(300,50)(320,50)\ArrowLine(320,50)(340,50)
\DashArrowArcn(270,50)(50,180,0){3}
\DashArrowArc(280,50)(20,0,90){3}\DashArrowArc(280,50)(20,90,180){3}
\Text(240,50)[]{\bf $\times$}\Text(280,70)[]{\bf $\times$}
\Text(200,45)[t]{$e_R$}\Text(340,45)[t]{$e_L$}
\Text(230,45)[t]{$\tilde{h}^0_1$}\Text(250,45)[t]{$\bar{\tilde{h}}{}^0_2$}
\Text(310,45)[t]{$\bar{\tilde{z}},\bar{\tilde{\gamma}}$}
\Text(260,67)[b]{$\tilde{t}_R$}\Text(300,67)[b]{$\tilde{t}_L$}
\Text(280,45)[t]{$t_L$}\Text(260,105)[b]{$\tilde{e}_L$}

\Text(260,0)[]{\bf (b)}

\end{picture}
\end{center}
\caption{Typical two-loop contributions to the electron EDM induced by
chargino and neutralino off-diagonal CP-violating wave-functions.
Insertion of the photon into all internal charged lines is
assumed.}\label{f4}
\end{figure}
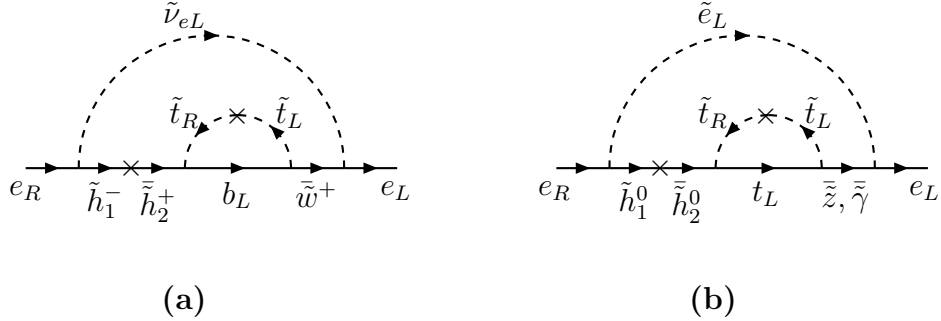

It  might now  seem  that  the CP-violating  phases  in the  effective
chargino-mass matrix  (\ref{Mchargino}) could almost be  absorbed by a
field redefinition  of the charginos.  However,  the same CP-violating
phases will  reappear in the interaction  Lagrangian (\ref{chqq}).  In
fact, CP-violating physical observables, such as the electron EDM, are
invariant  under  such  phase  rotations.   Therefore,  chargino-  and
neutralino-mediated diagrams such as those presented in Fig.\ \ref{f4}
are  expected to  give large  EDM effects, comparable to  the two-loop
Higgs-boson contributions discussed in \cite{CKP}.  An extensive study
of these novel EDM graphs will be given elsewhere \cite{APfuture}.

In summary, we have explicitly demonstrated that sizeable CP-violating
trilinear  couplings of  the Higgs  bosons to  third-generation scalar
quarks  can introduce  observable CP  violation into  the  gaugino and
higgsino sectors of the MSSM,  which is not attainable by an evolution
of the  SUSY RG  equations.  Based on  this minimal  CP-violating SUSY
framework,  we  have  shown  that  a new  Higgs-independent  class  of
two-loop   graphs  exists   which  may   lead  to   potentially  large
contributions  to the  electron  and neutron  EDM's.   Finally, it  is
important  to  stress  that  even  though the  two-loop  gaugino-  and
higgsino-mediated  contributions to  EDMs only  involve a  single SUSY
CP-violating  phase,  i.e.\ ${\rm  arg}\,  A_t$,  they  can result  in
different signs  related to  the signs of  the gaugino masses  and the
$\mu$ parameter,  and so allow for potential  cancellations with other
existing  third-generation scalar-quark  contributions studied  in the
literature \cite{DDLPD,CKP}.   The latter may  represent a technically
natural  as  well  as  phenomenologically appealing  solution  to  the
CP-crisis  problem in  SUSY theories.   In this  context, it  would be
interesting  to  analyze  in  detail the  phenomenological  impact  of
loop-induced gaugino/higgsino-sector CP violation on $B$-meson decays,
dark-matter searches and electroweak baryogenesis \cite{CPHig}.

\end{document}